\documentclass[aps,prl,twocolumn,groupedaddress]{revtex4}
\usepackage{graphicx}
\usepackage{amsmath}

\begin{document}

\title{An electromechanical Ising machine}

\author{I. Mahboob}

\email{imran.mahboob@lab.ntt.co.jp}

\author{H. Yamaguchi}

\affiliation{NTT Basic Research Laboratories, NTT Corporation, Atsugi-shi, Kanagawa 243-0198, Japan}

\begin{abstract}
Solving intractable mathematical problems in simulators composed of atoms, ions, photons or electrons has recently emerged as a subject of intense interest. Here we extend this concept to phonons that are localised in spectrally pure resonances in an electromechanical system which enables their interactions to be exquisitely fashioned via electrical means. We harness this platform to emulate the Ising Hamiltonian whose spin 1/2 particles are replicated by the phase bistable vibrations from a parametric resonance where multiple resonances play the role of a spin bath. The coupling between the mechanical pseudo spins is created by generating thermomechanical two-mode squeezed states which impart correlations between resonances that can imitate a ferromagnetic, random or an anti-ferromagnetic state on demand. These results suggest an electromechanical simulator for the Ising Hamiltonian could be built with a large number of spins with multiple degrees of coupling, a task that would overwhelm a conventional computer.
\end{abstract}

\maketitle

Electro/optomechanical systems consisting of a high quality mechanical resonance embedded in an electrical/optical transduction circuit \cite{roukes, ekinci, optomarq, cavopto} have emerged as versatile platform in which sensors with unprecedented sensitivities can be developed \cite{rugar2, NV, bach}, in which mechanical nonlinearities can be dynamically engineered and harvested \cite{mohanty, imran1, NL1} and even quantum mechanics within a macroscopic context can be studied \cite{gs, cav11, cav12}. However this unprecedented success has had the unfortunate consequence of stifling further advancement of mechanical systems beyond these extensively explored concepts. 

\begin{figure}[!hbt]
\vspace{0cm}
\includegraphics[scale=0.9]{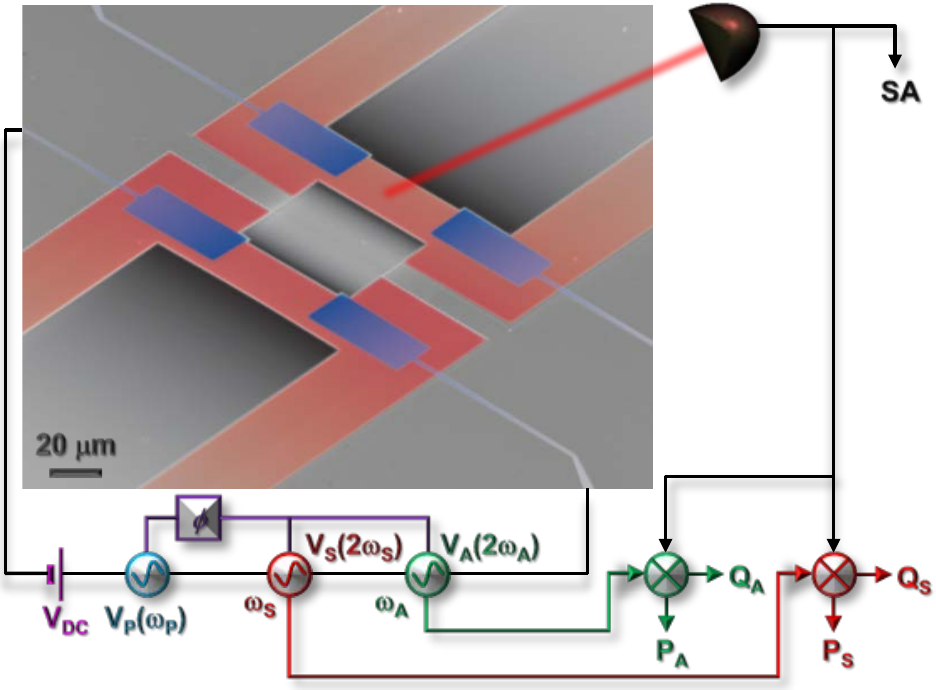}
\vspace{0.0cm}
\small{\caption{\textbf{The electromechanical system.} \textbf{a,} A false colour electron micrograph of the coupled mechanical resonators sustaining the symmetric and asymmetric vibration modes. The measurements were performed at room temperature and in a high vacuum and the mechanical vibrations were detected via a laser interferometer which was demodulated either in a spectrum analyser (SA) or in a phase sensitive detector utilising the heterodyne mixing setup detailed in the circuit schematic. Five signal generators were employed with two piezoelectrically activating the parametric resonances at $2\omega_n$, two generating the reference signals for the phase sensitive detectors at $\omega_n$ and the fifth creating the coupling between the parametric resonances via non-degenerate parametric down-conversion when activated at $\omega_{_\mathcal{P}}$.}}
\end{figure}

Meanwhile quantum simulators have been proposed as devices which could be used to solve problems in quantum physics that are beyond the capacity of classical computers \cite{iulia}. This notion can also be applied to mathematical problems for example the Ising Hamiltonian used to describe a spin glass which is characterised as an NP-hard problem that cannot be solved via conventional computing protocols \cite{Bara}. However in spite of this formidable challenge, the ground state spin configuration of the Ising Hamiltonian could yield solutions to optimisation problems that can be mapped on to its spin lattice and thus its efficient extraction is highly desired \cite{Kirk}. One approach to rapidly determining the ground state of the Ising Hamiltonian is to employ quantum annealing where quantum tunnelling is harnessed to search the energy landscape corresponding to a given problem programmed into the underlying spin configuration \cite{Ann1, Ann2, Ann3}. Recently an alternative and apparently classical approach to this problem has emerged where a time multiplexed optical parametric resonator network is programmed with an Ising Hamiltonian composed of four spins. The ground state is then determined by simply activating the network which preferentially resonates in its global potential minima \cite{Wang, Mara}. Here this concept is developed in a frequency multiplexed electromechanical parametric resonator and we show that this platform can readily be extended to multiple electromechanical parametric resonators with arbitrary degrees of coupling namely all the necessary prerequisites to solving the Ising Hamiltonian in a regime that has no analytical solution and where its numerical solution would demand exorbitant computational resources. 

The Ising model conceived using the tools of statistical physics describes ferromagnetism and in the absence of a magnetic field its Hamiltonian is given by $-\displaystyle\sum_{ik}^{_N}J_{ik}\sigma_i\sigma_k$ with $N$ particles each with two spin states $\sigma=\pm1$ and the coupling between the $i^{th}$ and $k^{th}$ particles is parametrised by $J_{ik}$. In the first steps to building an electromechanical Ising machine, the fundamentals of the Ising Hamiltonian need to be replicated in the electromechanical domain namely a mechanical pseudo spin to encode $\sigma$, multiple mechanical spins playing the role of an $N$ particle bath and finally coupling between the mechanical spins $J_{ik}$ whose magnitude and polarity can be controlled and has the potential to be extended beyond nearest neighbour spins.

The prototype electromechanical system in which these concepts are investigated is shown in Fig. 1 and it consists of two strongly coupled mechanical beams which sustain a symmetric (S) and an asymmetric (A) mode at $\omega_{_S}/2\pi \approx$245 kHz and $\omega_{_A}/2\pi \approx$260 kHz with quality factors ($\mathcal{Q}$) of 1300 and 2300 respectively (see Fig. S1 in Supplementary Information (SI)). The mechanical elements are integrated with piezoelectric transducers which enable the underlying harmonic potential of both modes to be parametrically modulated leading to a system Hamiltonian given by

\vspace{-0.0cm}
%\begin{center}
\begin{eqnarray}
H = \sum_{n=S}^A \Big(p_n^2/2 + \omega_n^2 q_n^2/2 \big(1-2\Gamma_n\cos(2\omega_nt) \nonumber \\ 
+ \beta_nq_n^2/2 \big)\Big) + \Lambda q_{_S} q_{_A}\cos(\omega_{_\mathcal{P}}t+\phi) 
\end{eqnarray}
%\end{center}   

\noindent where the summation expresses the kinetic and potential energies from both modes in terms of their position $q_n$ and momentum $p_n$ \cite{imran1, imSQUEE, dykman}. The potential energy term contains three contributions, the second arising from the periodic modulation of the mechanical spring constant with amplitude $\Gamma_n$ at twice the natural mode frequency which yields degenerate parametric amplification and parametric resonance \cite{rugars, five} and the third due to the well-known anharmonicity from the Duffing nonlinearity $\beta_n$ which appears at large displacements \cite{clee}. A unique feature of the parametric resonance is that it can only vibrate with two phases and thus it provides the ideal construct within which a classical spin can be harboured \cite{Wang, imran1}. The last term describes non-degenerate parametric down-conversion when the periodic modulation is activated with a pump ($\mathcal{P}$) amplitude $\Lambda$ at the sum frequency of both modes ($\omega_{_\mathcal{P}} = \omega_{_S} + \omega_{_A}$) which results in the symmetric and asymmetric modes becoming amplified and correlated yielding a two-mode squeezed state \cite{imSQUEE}. 

\begin{figure}[!ht]
\vspace{0cm}
\includegraphics[scale=0.9]{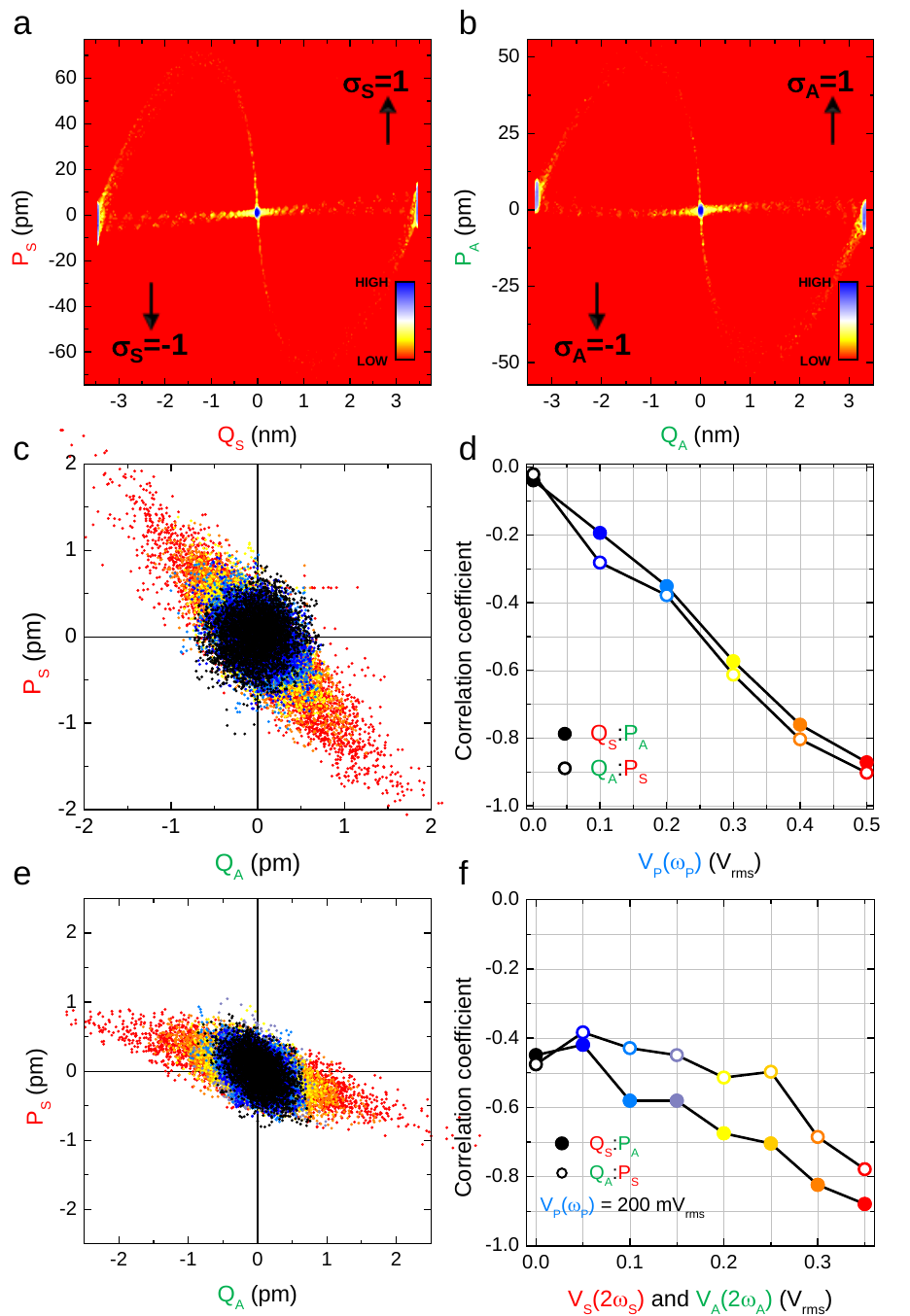}
\vspace{0.0cm}
\small{\caption{\textbf{Mechanical spins and spin coupling.} \textbf{a} and \textbf{b,}  The occupation probabilities of the parametric resonances in phase space are measured with the parametric actuation periodically activated with $V_{_S}(2\omega_{_S})$ = 0.7 V$_{rms}$ and  $V_{_A}(2\omega_{_A})$ = 0.6 V$_{rms}$ respectively. The resultant phase portraits indicate that when a parametric resonance is activated, each mode evolves from its stationary state at the origin, which is broadened by thermomechanical noise fluctuations, to one of the two available oscillating states with a $\pi$ phase separation that can be exploited to host classical spin information. \textbf{c,} The phase portrait reconstructed from the cross-quadratures of the symmetric and asymmetric mode as function of pump intensity. \textbf{d,} The corresponding correlation coefficient reveals that the two modes become indistinguishable as the pump amplitude is increased where the colours indicate the amplitudes in the previous figure. \textbf{e,} A two-mode squeezed state generated with $V_{_\mathcal{P}}(2\omega_{_\mathcal{P}})$ = 0.2 V$_{rms}$ which is then enhanced via degenerate parametric amplification of both modes. \textbf{f,} The corresponding correlation coefficient confirms that the initial weakly correlated state can be enhanced by the degenerate parametric amplification where the colour coding indicates the amplitudes in the previous figure.}}
\end{figure}

This Hamiltonian can be transformed in the rotating frame approximation with the introduction of a dimensionless canonical position coordinate $Q_n$ and conjugate momentum $P_n$, as detailed in SI, which yields

\vspace{-0.0cm}
%\begin{center}
\begin{eqnarray}
\mathcal{H} = -\sum_{n=S}^A \big(\omega_n\Gamma_n^2/3\beta_n^2 \big) + J_{_{SA}}\sigma_{_S}\sigma_{_A} 
\end{eqnarray}
%\end{center} 

\noindent where the first term describes the energy corresponding to the two stable oscillation phases of a parametric resonance at $P_n=0$ and $Q_n=\sigma_n\sqrt{4\Gamma_n/3\beta_n}$ \cite{dykman, imCHA} and the second term quantifies the coupling $J_{_{SA}}=\Lambda \cos(\phi)\sqrt{\Gamma_{_S}\Gamma_{_A}/9\omega_{_S}\omega_{_A}\beta_{_S}\beta_{_A}}$ between the modes with their two phases encoding $\sigma_n=\pm1$. Thus this electromechanical system can be formally mapped onto the Ising model composed of two classical spins arising from the phase bistable parametric resonances of both modes that can be coupled from the non-degenerate parametric down-conversion. Pleasingly, this analysis also reveals that the magnitude of the coupling between the mechanical spins created by $\Lambda$ can be enhanced by the degenerate parametric resonances that are activated when $\Gamma_n \ge \gamma_n$, where the damping rate $\gamma_n=\omega_n/\mathcal{Q}_n$ \cite{imran1}, and its polarity i.e. the sign of $J_{_{SA}}$ can be selected by simply tuning the pump phase $\phi$ \cite{sqz8}.

\begin{figure*}[!ht]
\vspace{0cm}
\includegraphics[scale=0.9]{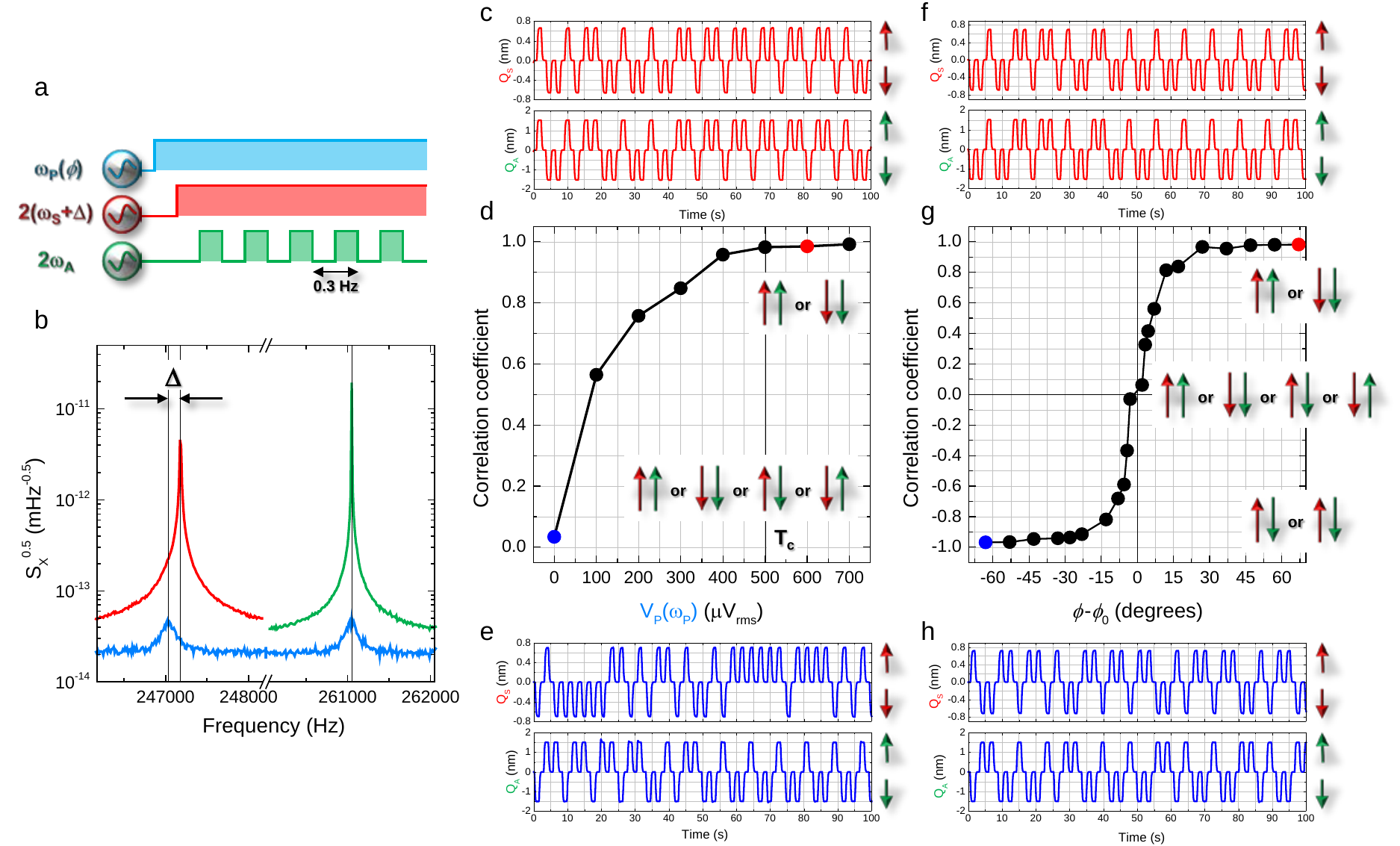}
\vspace{0cm}
\small{\caption{\textbf{Emulating the Ising model in the mechanical domain.} \textbf{a,} The pulse sequence used to demonstrate the fundamentals of the Ising Hamiltonian in an electromechanical system. \textbf{b,} The spectral response of the symmetric and asymmetric modes in response to the above pulse sequence. \textbf{c,} The temporal response of the mechanical spins to the above pulse sequence which are readout via the in-phase component of their motion with $V_{_S}(2(\omega_{_S}+\Delta))$ = 0.7 V$_{rms}$, $V_{_A}(2\omega_{_A})$ = 0.7 V$_{rms}$ and $V_{_\mathcal{P}}(\omega_{_\mathcal{P}})$=600 $\mu$V$_{rms}$ which reveals ferromagnetic ordering with the two spins always in parallel alignment where the spin orientations are visualised by the arrows with red (green) indicating the symmetric (asymmetric) mode. All the temporal outputs are measured over 1000 seconds to enable statistically reliable analysis. \textbf{d,} The correlation coefficient extracted from the switching outputs as a function of pump intensity. At the largest pump amplitude the mechanical spins exhibit ferromagnetic ordering but as the pump is reduced to zero, the spins become disordered undergoing an effective second order phase transition with the pump playing the role of temperature and a critical temperature (T$_c$) of 500 $\mu$V$_{rms}$. \textbf{e,} A segment of the temporal output used to extract the above correlation coefficient with the pump deactivated which clearly reveals that the mechanical spins have become decoupled yielding uncorrelated switching events. \textbf{g,} The correlation coefficient extracted from the mechanical system in response to the above pulse sequence as a function of the pump phase with $V_{_\mathcal{P}}(\omega_{_\mathcal{P}})$=600 $\mu$V$_{rms}$. As the pump phase is adjusted, the mechanical spins exhibiting ferromagnetic ordering at $\phi-\phi_0$=67$^{\circ}$ detailed in the temporal output in \textbf{f} transition through a disordered state with random spin alignment to finally an anti-correlated state at $\phi-\phi_0$=-63$^{\circ}$ with the spins exhibiting anti-ferromagnetic ordering as detailed in the temporal output in \textbf{h}. Note that the value of $\phi_0$ is determined by the spectral position of the readout probe with respect to the resonances of the two modes. }}
\end{figure*}

To experimentally confirm these expectations, the spring constant of either mode is modulated at $V_n(2\omega_n)$ via the stress induced by the piezoelectric transducers namely $\Gamma_n$ in the above equations is activated \cite{imran1}. This results in a given modes thermomechanical fluctuations initially being enhanced corresponding to degenerate parametric amplification \cite{rugars} and once the rate at which the phonons are generated begins to exceed their rate of decay namely $\Gamma_n \ge \gamma_n$ it results in infinite amplification corresponding to a parametric resonance as depicted in Figs. S2 \cite{five, imran1}. The parametric resonance, under periodic driving, is then projected into phase space by mixing with a local oscillator locked onto its resonance frequency and demodulated in a phase sensitive detector (see circuit schematic in Fig. 1). This measurement yields two time series for the in-phase $Q_n$ and quadrature $P_n$ components that enables a phase portrait to be reconstructed as shown in Figs. 2a and 2b which probabilistically unveils the parametric resonances from both modes being able to vibrate with only two phases separated by $\pi$ radians as described above \cite{imran1}. This phase bistable vibration enables a classical spin to be mimicked in the mechanical domain where hence forth the positive in-phase component is defined as spin up i.e. $\sigma_n=1$ and vice versa.

In order to create coupling between the pseudo spins encapsulated by the symmetric and asymmetric modes, a mechanical two-mode squeezed state is created by pumping the spring constant of both modes at $V_{_\mathcal{P}}(\omega_{_\mathcal{P}})$ corresponding to $\Lambda$ in the above equations \cite{imSQUEE}. This results in the thermomechanical fluctuations of both modes being simultaneously amplified from non-degenerate parametric down-conversion as shown in Fig. S2 and projecting this output in phase space reveals it to be phase insensitive as shown Fig. S4 \cite{sqz07}. In contrast, the degenerate parametric amplification detailed above is phase sensitive with only the in-phase component being amplified whilst the quadrature component is squeezed as shown in Fig. S3 \cite{rugars}. However, the simultaneous generation of phonons in both modes also results in their vibrations becoming correlated which can be identified by reconstructing their cross-quadratures in phase space namely the in-phase component of the asymmetric mode versus the quadrature component of the symmetric mode (or vice versa) as shown in Fig. 2c (Fig. S4c). The resultant squashed distribution implies that the motion from both modes has become intertwined which can statistically be confirmed by evaluating their correlation coefficient $cov(Q_{_A}P_{_S})/\zeta_{Q_{_A}}\zeta_{P_{_S}}$ where the numerator describes the covariance and $\zeta$ is the standard deviation. The results of this analysis (also including $Q_{_S}:P_{_A}$ whose phase portraits are shown in Fig. S4) as a function of pump intensity are shown in Fig. 2d which reveals that in this process the correlation coefficient tends to -1 indicating a perfect anti-correlation that is phonons pairs are simultaneously generated in both modes and with an increasing rate which renders their vibrations classically indistinguishable.  

To transfer this correlation into the parametric resonances, a weakly correlated state is first created at $V_{_\mathcal{P}}(\omega_{_\mathcal{P}})$=200 mV$_{rms}$ as shown in Figs. 2e and 2f. Next degenerate parametric amplification is continuously activated in both modes and the resultant correlation coefficient is monitored as function of both $V_{_S}(2\omega_{_S})$ and $V_{_A}(2\omega_{_A})$. As expected from equation 2, the correlations between the two modes can be enhanced from the degenerate parametric amplification which can readily be seen by the narrower squashed distribution in phase space as shown in Fig. 2e and quantitatively confirmed via the correlation coefficient in Fig. 2f. Indeed once maximum degenerate parametric amplification has been achieved and both modes start to parametrically resonate, a perfectly correlated state is created even with weak pump intensity. Although experimentally the evolution of the correlation coefficient across the transition from degenerate parametric amplification to parametric resonance cannot be seamlessly measured as the phase sensitive detectors became overloaded, this transition is readily confirmed both numerically and analytically by solving equation 1 in this configuration as shown in Figs. S5 and S6.    

Based on the above results, the essential features of the Ising Hamiltonian between two spin 1/2 particles in the electromechanical domain can now be demonstrated using the pulse sequence depicted in Fig. 3a. First a two-mode squeezed state is created which is undetectable via thermomechanical fluctuations namely $V_{_\mathcal{P}}(\omega_{_\mathcal{P}}) \rightarrow 0$. Next, the symmetric mode is activated off-resonance i.e. $2(\omega_{_S}+\Delta)$ which results in no vibration. This is then followed by the asymmetric mode being parametrically activated at $2\omega_{_A}$ where the tension created by its motion modifies the spring constant and shifts the natural frequency of the symmetric mode by $\Delta$ resulting in its parametric resonance being simultaneously activated as shown in Fig. 3b. In this formation, periodically triggering the asymmetric mode results in the symmetric mode becoming activating with the same period and monitoring the in-phase component from both modes on resonance enables identification of their spin orientation (see Figs. 2a and 2b). Correlations from the two-mode squeezing can then be statistically quantified by monitoring the resultant switching outputs from both modes. For instance, activating the asymmetric mode will result in either a spin up or a spin down the choice of which is determined by its thermal fluctuations. However, if a two-mode squeezed state is created then the switching output observed from the symmetric mode will be correlated with that from the asymmetric mode.

Implementing this protocol with a weak pump yields the output shown in Fig. 3c where an up spin from the asymmetric mode coincides with an up spin in the symmetric mode and vice versa. However if the pump is deactivated, the two modes become uncoupled and switch randomly as shown in Fig. 3e. The correlation coefficient can then be extracted from these switching outputs as a function of pump intensity as depicted in Fig. 3d. This reveals the mechanical pseudo spins can achieve parallel alignment when the pump is activated and the corresponding correlation coefficient tends to unity. This parallel configuration can be controllably deactivated via the pump and the correlation coefficient smoothly transitions to zero mimicking an effective second order ferromagnetic phase transition. Consequently $V_{_\mathcal{P}}(\omega_{_\mathcal{P}})$ can convincingly play the role of $J_{ik}$ in the Ising Hamiltonian.

Now in order to control the sign of the coupling, the pump phase is adjusted as depicted in Fig. 1 and detailed in equation 2. Indeed measuring the two-mode squeezing via thermomechanical noise fluctuations in phase-space as a function of $\phi$ clearly reveals that the anti-correlated state depicted in Fig. 2c can be tuned to a correlated state as shown in Fig. S7 \cite{sqz8}. Repeating the protocol outlined in Fig. 3a but now as function of $\phi$ with a sufficiently intense pump so that correlations are generated with perfect fidelity yields the result shown in Fig. 3g. Here the parallel spin alignment from the two modes as captured in Fig. 3f can be continuously tuned to an anti-parallel alignment via $\phi$ as shown in Fig. 3h corresponding to an effective phase transition from ferromagnetism to anti-ferromagnetism or in other words $J_{ik} \rightarrow -J_{ik}$ in the Ising model.

Indeed the pulse sequence shown in Fig. 3a can also be executed with the off resonant excitation of the asymmetric mode in the second step and triggering via the symmetric mode as detailed in Fig. S8. Interestingly, the data shown in Fig. 3c (and Fig. S8c) indicates that correlations from the two-mode squeezing can be observed via the parametric resonances with a pump amplitude that is three orders of magnitude smaller than the equivalent measurement utilising thermomechanical noise fluctuations as shown in Fig. 2d. This observation suggests an alternative approach to identifying mechanical vacuum squeezed states via the large signal from their parametric resonance instead of their vacuum noise \cite{TYam} which could prove pivotal in detecting a macroscopic all-mechanical entangled state \cite{imSQUEE}.  

The results detailed in Fig. 3 confirm that the requisite features of the Ising Hamiltonian can be reproduced in the electromechanical domain. However its implementation with two mechanical modes in this instance is trivial and therefore it is instructive to examine the prospects of an electromechanical Ising machine composed of a large number of spin particles with multiple degrees of coupling.

\begin{figure}[!ht]
\vspace{0cm}
\includegraphics[scale=1.0]{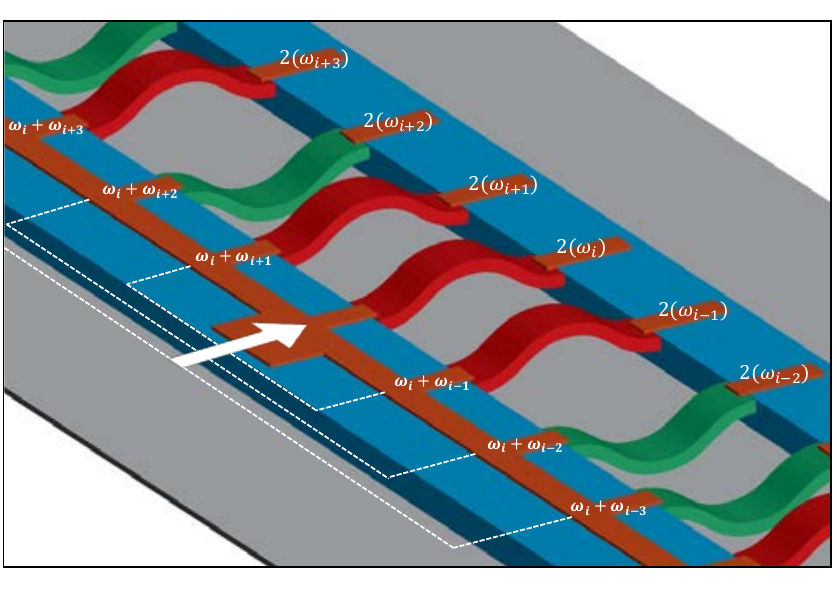}
\vspace{0cm}
\small{\caption{\textbf{The electromechanical Ising machine.} A conceptual image of an electromechanical Ising machine based on an array of mechanical elements each parametrically resonating to encode classical spin information (red corresponds to spin up and green to spin down) where two-mode squeezing is used to create coupling between elements. The parametric resonances are piezoelectrically activated and readout via the individual gate electrodes (orange) on each mechanical element where the $i^{th}$ element has a natural frequency $\omega_i$. The global gate electrode (orange) located on the left clamping point of all the mechanical elements can enable coupling between any pair of mechanical spins with the application of a sum frequency pump modulation. Utilising FDM, the pump can execute multiple degrees of coupling between a large number of spins potentially enabling the Ising Hamiltonian to be solved in a configuration that is beyond the reach of conventional computers.}}
\end{figure}

Based on the above demonstrations, an electromechanical Ising machine visualised in Fig. 4 would seem feasible. Here multiple mechanical elements with different frequencies are weakly mechanically coupled to their neighbours. The elements encode spin information via the bistable phase of their piezoelectrically activated parametric resonance in their fundamental mode, from the right clamping point, where this information can also be pre-programmed \cite{imran1}. The key difference here is that spin information is stored in each mechanical element rather than modes composed of more than one element as demonstrated above. On the left clamping point a {\it coupling} gate electrode is defined which interconnects the mechanical spins via the piezoelectrically activated parametric down-conversion pump at the sum frequency of two elements for instance $\omega_i+\omega_{i+1}$ as shown in Fig. 4. Naturally, the reduced mechanical coupling between the elements will require a stronger pump amplitude which would be feasible as even weak pumping can efficiently couple the mechanical spins as demonstrated in Fig. 3d. Each mechanical spin could then be readily coupled to all its neighbours by employing a frequency division multiplexed (FDM) pump composed of multiple sum frequencies where Fig. 4 explicitly depicts the FDM pump input necessary to couple the $i^{th}$ element to its first six nearest neighbours. The compact form of this coupling is in stark contrast to the optical Ising machine which requires delay lines that increase both in number for more spins and length for higher order couplings \cite{Mara}. The FDM piezoelectric pump in principle enables a large number of mechanical spins to sustain multiple degrees of coupling permitting the electromechanical Ising machine to be programmed with problems that are unsolvable with conventional computers. Although this platform could tackle NP-hard problems \cite{Wang, Mara}, it does not offer speed up as it employs classical annealing where thermomechanical fluctuations of the mechanical elements drive the search in the underlying potential energy landscape for the ground state. However tantalisingly if all the mechanical elements are operated in their ground state \cite{gs, cav11, cav12} then quantum tunnelling could be harnessed to increase the speed of search thus making an inexorable case for development of such an electromechanical Ising machine.

\vspace{0.2cm}

\vspace{0.2cm}

\vspace{0.2cm}

%\noindent{\bf Methods}

%\bibliography{scibib}

\begin{thebibliography}{10}
\expandafter\ifx\csname url\endcsname\relax
  \def\url#1{\texttt{#1}}\fi
\expandafter\ifx\csname urlprefix\endcsname\relax\def\urlprefix{URL }\fi
\providecommand{\bibinfo}[2]{#2}
\providecommand{\eprint}[2][]{\url{#2}}

\bibitem{roukes}
\bibinfo{author}{Roukes, M.}
\newblock \bibinfo{title}{Nanoelectromechanical systems face the future}.
\newblock \emph{\bibinfo{journal}{Phys. World}} \textbf{\bibinfo{volume}{14}},
  \bibinfo{pages}{25--31} (\bibinfo{year}{2001}).

\bibitem{ekinci}
\bibinfo{author}{Ekinci, K.~L.} \& \bibinfo{author}{Roukes, M.~L.}
\newblock \bibinfo{title}{Nanoelectromechanical systems}.
\newblock \emph{\bibinfo{journal}{Rev. Sci. Instrum.}}
  \textbf{\bibinfo{volume}{76}}, \bibinfo{pages}{061101}
  (\bibinfo{year}{2005}).

\bibitem{optomarq}
\bibinfo{author}{Marquardt, F.} \& \bibinfo{author}{Girvin, S.~M.}
\newblock \bibinfo{title}{Optomechanics}.
\newblock \emph{\bibinfo{journal}{Physics}} \textbf{\bibinfo{volume}{2}},
  \bibinfo{pages}{40} (\bibinfo{year}{2009}).

\bibitem{cavopto}
\bibinfo{author}{Aspelmeyer, M.}, \bibinfo{author}{Kippenberg, T.~J.} \&
  \bibinfo{author}{Marquardt, F.}
\newblock \bibinfo{title}{Cavity optomechanics}.
\newblock \emph{\bibinfo{journal}{Rev. Mod. Phys.}}
  \textbf{\bibinfo{volume}{86}}, \bibinfo{pages}{1391} (\bibinfo{year}{2014}).

\bibitem{rugar2}
\bibinfo{author}{Rugar, D.}, \bibinfo{author}{Budakian, R.},
  \bibinfo{author}{Mamin, H.~J.} \& \bibinfo{author}{Chui, B.~W.}
\newblock \bibinfo{title}{Single spin detection by magnetic resonance force
  microscopy}.
\newblock \emph{\bibinfo{journal}{Nature}} \textbf{\bibinfo{volume}{430}},
  \bibinfo{pages}{329} (\bibinfo{year}{2004}).

\bibitem{NV}
\bibinfo{author}{Arcizet, O.} \emph{et~al.}
\newblock \bibinfo{title}{A single nitrogen-vacancy defect coupled to a
  nanomechanical oscillator}.
\newblock \emph{\bibinfo{journal}{Nature Phys.}} \textbf{\bibinfo{volume}{7}},
  \bibinfo{pages}{879--883} (\bibinfo{year}{2011}).

\bibitem{bach}
\bibinfo{author}{Chaste, J.} \emph{et~al.}
\newblock \bibinfo{title}{A nanomechanical mass sensor with yoctogram
  resolution}.
\newblock \emph{\bibinfo{journal}{Nature Nanotech.}}
  \textbf{\bibinfo{volume}{7}}, \bibinfo{pages}{301--304}
  (\bibinfo{year}{2012}).

\bibitem{mohanty}
\bibinfo{author}{Badzey, R.~L.} \& \bibinfo{author}{Mohanty, P.}
\newblock \bibinfo{title}{Coherent signal amplification in bistable
  nanomechanical oscillators by stochastic resonance}.
\newblock \emph{\bibinfo{journal}{Nature}} \textbf{\bibinfo{volume}{437}},
  \bibinfo{pages}{995} (\bibinfo{year}{2005}).

\bibitem{imran1}
\bibinfo{author}{Mahboob, I.} \& \bibinfo{author}{Yamaguchi, H.}
\newblock \bibinfo{title}{Bit storage and bit flip operations in an
  electromechanical oscillator}.
\newblock \emph{\bibinfo{journal}{Nature Nanotech.}}
  \textbf{\bibinfo{volume}{3}}, \bibinfo{pages}{275--279}
  (\bibinfo{year}{2008}).

\bibitem{NL1}
\bibinfo{author}{Eichler, A.} \emph{et~al.}
\newblock \bibinfo{title}{Nonlinear damping in mechanical resonators made from
  carbon nanotubes and graphene}.
\newblock \emph{\bibinfo{journal}{Nature Nanotech.}}
  \textbf{\bibinfo{volume}{6}}, \bibinfo{pages}{339--342}
  (\bibinfo{year}{2011}).

\bibitem{gs}
\bibinfo{author}{O'Connell, A.~D.} \emph{et~al.}
\newblock \bibinfo{title}{Quantum ground state and single-phonon control of a
  mechanical resonator}.
\newblock \emph{\bibinfo{journal}{Nature}} \textbf{\bibinfo{volume}{464}},
  \bibinfo{pages}{697--703} (\bibinfo{year}{2010}).

\bibitem{cav11}
\bibinfo{author}{Teufel, J.~D.} \emph{et~al.}
\newblock \bibinfo{title}{Sideband cooling of micromechanical motion to the
  quantum ground state}.
\newblock \emph{\bibinfo{journal}{Nature}} \textbf{\bibinfo{volume}{475}},
  \bibinfo{pages}{359--Â–363} (\bibinfo{year}{2011}).

\bibitem{cav12}
\bibinfo{author}{Chan, J.} \emph{et~al.}
\newblock \bibinfo{title}{Laser cooling of a nanomechanical oscillator into its
  quantum ground state}.
\newblock \emph{\bibinfo{journal}{Nature}} \textbf{\bibinfo{volume}{478}},
  \bibinfo{pages}{89--Â–92} (\bibinfo{year}{2011}).

\bibitem{iulia}
\bibinfo{author}{Buluta, I.} \& \bibinfo{author}{Nori, F.}
\newblock \bibinfo{title}{Quantum simulators}.
\newblock \emph{\bibinfo{journal}{Science}} \textbf{\bibinfo{volume}{326}},
  \bibinfo{pages}{108--111} (\bibinfo{year}{2009}).

\bibitem{Bara}
\bibinfo{author}{Barahona, F.}
\newblock \bibinfo{title}{On the computational complexity of ising spin glass
  models}.
\newblock \emph{\bibinfo{journal}{J. Phys. A: Math. Gen.}}
  \textbf{\bibinfo{volume}{15}}, \bibinfo{pages}{3241} (\bibinfo{year}{1982}).

\bibitem{Kirk}
\bibinfo{author}{Kirkpatrick, S.}, \bibinfo{author}{Jr, C. D.~G.} \&
  \bibinfo{author}{Vecchi, M.~P.}
\newblock \bibinfo{title}{Optimization by simulated annealing}.
\newblock \emph{\bibinfo{journal}{Science}} \textbf{\bibinfo{volume}{220}},
  \bibinfo{pages}{671--680} (\bibinfo{year}{1983}).

\bibitem{Ann1}
\bibinfo{author}{Finnila, A.~B.}, \bibinfo{author}{Gomez, M.~A.},
  \bibinfo{author}{Sebenik, C.}, \bibinfo{author}{Stenson, C.} \&
  \bibinfo{author}{Doll, J.~D.}
\newblock \bibinfo{title}{Quantum annealing: A new method for minimizing
  multidimensional functions}.
\newblock \emph{\bibinfo{journal}{Chem. Phys. Lett.}}
  \textbf{\bibinfo{volume}{219}}, \bibinfo{pages}{343--348}
  (\bibinfo{year}{1994}).

\bibitem{Ann2}
\bibinfo{author}{Santoro, G.~E.}, \bibinfo{author}{Marto\^{n}\'{a}k, R.},
  \bibinfo{author}{Tosatti, E.} \& \bibinfo{author}{Car, R.}
\newblock \bibinfo{title}{Theory of quantum annealing of an ising spin glass}.
\newblock \emph{\bibinfo{journal}{Science}} \textbf{\bibinfo{volume}{292}},
  \bibinfo{pages}{472--475} (\bibinfo{year}{2002}).

\bibitem{Ann3}
\bibinfo{author}{Johnson, M.~W.} \emph{et~al.}
\newblock \bibinfo{title}{Quantum annealing with manufactured spins}.
\newblock \emph{\bibinfo{journal}{Nature}} \textbf{\bibinfo{volume}{473}},
  \bibinfo{pages}{194--198} (\bibinfo{year}{2011}).

\bibitem{Wang}
\bibinfo{author}{Wang, Z.}, \bibinfo{author}{Marandi, A.},
  \bibinfo{author}{Wen, K.}, \bibinfo{author}{Byer, R.~L.} \&
  \bibinfo{author}{Yamamoto, Y.}
\newblock \bibinfo{title}{Coherent ising machine based on degenerate optical
  parametric oscillators}.
\newblock \emph{\bibinfo{journal}{Phys. Rev. A}} \textbf{\bibinfo{volume}{88}},
  \bibinfo{pages}{063853} (\bibinfo{year}{2013}).

\bibitem{Mara}
\bibinfo{author}{Marandi, A.}, \bibinfo{author}{Wang, Z.},
  \bibinfo{author}{Takata, K.}, \bibinfo{author}{Byer, R.~L.} \&
  \bibinfo{author}{Yamamoto, Y.}
\newblock \bibinfo{title}{Network of time-multiplexed optical parametric
  oscillators as a coherent ising machine}.
\newblock \emph{\bibinfo{journal}{Nature Photon.}}
  \textbf{\bibinfo{volume}{8}}, \bibinfo{pages}{937--942}
  (\bibinfo{year}{2014}).

\bibitem{imSQUEE}
\bibinfo{author}{Mahboob, I.}, \bibinfo{author}{Okamoto, H.},
  \bibinfo{author}{Onomitsu, K.} \& \bibinfo{author}{Yamaguchi, H.}
\newblock \bibinfo{title}{Two-mode thermal-noise squeezing in an
  electromechanical resonator}.
\newblock \emph{\bibinfo{journal}{Phys. Rev. Lett.}}
  \textbf{\bibinfo{volume}{113}}, \bibinfo{pages}{167203}
  (\bibinfo{year}{2014}).

\bibitem{dykman}
\bibinfo{author}{Ryvkine, D.} \& \bibinfo{author}{Dykman, M.~I.}
\newblock \bibinfo{title}{Resonant symmetry lifting in a parametrically
  modulated oscillator}.
\newblock \emph{\bibinfo{journal}{Phys. Rev. E}} \textbf{\bibinfo{volume}{74}},
  \bibinfo{pages}{061118} (\bibinfo{year}{2006}).

\bibitem{rugars}
\bibinfo{author}{Rugar, D.} \& \bibinfo{author}{Gr\"{u}tter, P.}
\newblock \bibinfo{title}{Mechanical parametric amplification and
  thermomechanical noise squeezing}.
\newblock \emph{\bibinfo{journal}{Phys. Rev. Lett.}}
  \textbf{\bibinfo{volume}{67}}, \bibinfo{pages}{699--702}
  (\bibinfo{year}{1991}).

\bibitem{five}
\bibinfo{author}{Turner, K.~L.} \emph{et~al.}
\newblock \bibinfo{title}{Five parametric resonances in a
  microelectromechanical system}.
\newblock \emph{\bibinfo{journal}{Nature}} \textbf{\bibinfo{volume}{396}},
  \bibinfo{pages}{149} (\bibinfo{year}{1998}).

\bibitem{clee}
\bibinfo{author}{Aldridge, J.~S.} \& \bibinfo{author}{Cleland, A.~N.}
\newblock \bibinfo{title}{Noise-enabled precision measurements of a duffing
  nanomechanical resonator}.
\newblock \emph{\bibinfo{journal}{Phys. Rev. Lett.}}
  \textbf{\bibinfo{volume}{94}}, \bibinfo{pages}{156403}
  (\bibinfo{year}{2005}).

\bibitem{imCHA}
\bibinfo{author}{Mahboob, I.}, \bibinfo{author}{Froitier, C.} \&
  \bibinfo{author}{Yamaguchi, H.}
\newblock \bibinfo{title}{A symmetry-breaking electromechanical detector}.
\newblock \emph{\bibinfo{journal}{Appl. Phys. Lett.}}
  \textbf{\bibinfo{volume}{96}}, \bibinfo{pages}{213103}
  (\bibinfo{year}{2010}).

\bibitem{sqz8}
\bibinfo{author}{Flurin, E.}, \bibinfo{author}{Roch, N.},
  \bibinfo{author}{Mallet, F.}, \bibinfo{author}{Devoret, M.~H.} \&
  \bibinfo{author}{Huard, B.}
\newblock \bibinfo{title}{Generating entangled microwave radiation over two
  transmission lines}.
\newblock \emph{\bibinfo{journal}{Phys. Rev. Lett.}}
  \textbf{\bibinfo{volume}{109}}, \bibinfo{pages}{183901}
  (\bibinfo{year}{2012}).

\bibitem{sqz07}
\bibinfo{author}{Bergeal, N.} \emph{et~al.}
\newblock \bibinfo{title}{Phase-preserving amplification near the quantum limit
  with a josephson ring modulator}.
\newblock \emph{\bibinfo{journal}{Nature}} \textbf{\bibinfo{volume}{465}},
  \bibinfo{pages}{64--68} (\bibinfo{year}{2010}).

\bibitem{TYam}
\bibinfo{author}{Lin, Z.~R.} \emph{et~al.}
\newblock \bibinfo{title}{Josephson parametric phase-locked oscillator and its
  application to dispersive readout of superconducting qubits}.
\newblock \emph{\bibinfo{journal}{Nature Commun.}}
  \textbf{\bibinfo{volume}{5}}, \bibinfo{pages}{4480} (\bibinfo{year}{2014}).

\end{thebibliography}

\vspace{0.5cm}

\small{

\vspace{0.5cm}

\noindent \textbf{Acknowledgements} The authors are grateful to H. Okamoto, N. Kitajima for fabricating the device to Y. Ishikawa, K. Onomitsu for growing the heterostructure and to N. Lambert, F. Nori for discussions and comments. This work was partly supported by JSPS KAKENHI Grant No. 23241046. 

\vspace{0.5cm}

\noindent \textbf{Author contributions} H.Y. and I.M. conceived the idea, I.M. performed the measurements, analysed the data, conducted the numerical modelling and wrote the paper. H.Y. developed the analytical model and planned the project. 

\vspace{0.5cm}

\noindent \textbf{Supplementary Information} is linked to the online version of the paper.

\vspace{0.5cm}

\noindent \textbf{Competing financial interests} The authors declare that they have no competing financial interests.}

\end{document}